\documentclass[lettersize,journal]{IEEEtran}
\usepackage{amsmath,amsfonts}
\usepackage{algorithmic}
\usepackage{algorithm}
\usepackage{array}
\usepackage[caption=false,font=normalsize,labelfont=sf,textfont=sf]{subfig}
\usepackage{textcomp}
\usepackage{stfloats}
\usepackage{url}
\usepackage{cite}
\usepackage{verbatim}
\usepackage{graphicx}
\usepackage{setspace}
\usepackage{booktabs}
\usepackage{enumerate}
\usepackage{diagbox}
\usepackage{arydshln}
\usepackage{accents}
\usepackage{bm}
\usepackage{balance}

\def\BibTeX{{\rm B\kern-.05em{\sc i\kern-.025em b}\kern-.08em
    T\kern-.1667em\lower.7ex\hbox{E}\kern-.125emX}}
\usepackage{balance}
\begin{document}
\title{{A Novel Block-Wise Index Modulation Scheme for High-Mobility OTFS Communications}}
\author{\small{
		\IEEEauthorblockN{Mi Qian\IEEEauthorrefmark{1}, Yao Ge\IEEEauthorrefmark{2}, Miaowen Wen\IEEEauthorrefmark{1}, and Fei Ji\IEEEauthorrefmark{1} }

		\IEEEauthorblockA{\textit{\IEEEauthorrefmark{1}School of Electronics and Information Engineering, South China University of Technology, Guangzhou 510641, China}\\
			\textit{\IEEEauthorrefmark{2}Continental-NTU Corporate Lab, Nanyang Technological University, 639798, Singapore}\\
			Email: eemqian@mail.scut.edu.cn, yao.ge@ntu.edu.sg, \{eemwwen, eefeiji\}@scut.edu.cn
		}\vspace{-2em}
	}
\thanks{This work was supported in part by the National Natural Science Foundation of China under Grant 61871190, and in part by the Guangdong Basic and Applied Basic Research Project under Grant 2021B1515120067.}
}


\maketitle

\begin{abstract}
As a promising technique for high-mobility wireless communications,  
orthogonal time frequency space (OTFS) has been proved to enjoy excellent advantages with respect to traditional orthogonal frequency division multiplexing (OFDM). However, a challenging problem is to design efficient systems to further improve the performance. In this paper, we propose a novel block-wise index modulation (IM) scheme for OTFS systems, named Doppler-IM with OTFS (DoIM-OTFS), where a block of Doppler resource bins are activated simultaneously. For practical implementation, we develop a low complexity customized message passing (CMP) algorithm for our proposed DoIM-OTFS scheme. Simulation results demonstrate our proposed DoIM-OTFS system outperforms traditional OTFS system without IM. The proposed CMP algorithm can achieve desired performance and robustness to the imperfect channel state information (CSI).
\end{abstract}


\section{Introduction}
Nowadays, a large number of wireless applications such as high-speed trains and unmanned autonomous vehicles are emerging. Accordingly, it is imperative to have a high data rate and low latency communications to satisfy the fast-growing requirements in the future. OFDM modulation has been prevailing so far as it is able to provide high-spectrum efficiency and is easy to implement \cite{7936676,6587554}. However, for  time-varying channels with large Doppler spread, OFDM will suffer from terrible performance degradation due to the loss of orthogonality or inter-carrier-interference~(ICI). 

To cope with the high-mobility scenarios, a new modulation scheme referred to as OTFS has been proposed \cite{7925924}, which can achieve significant performance improvement than OFDM modulation. OTFS can exploit the diversity gain in both the delay and Doppler dimensions of a mobile wireless channel since all the transmitted symbols can be multiplexed in the delay-Doppler domain and spread over the time-frequency domain \cite{8424569,9557830}. Furthermore, OTFS can convert the time-varying channel into a two-dimensional (2D) quasi-time-invariant channel in the delay-Doppler domain, which significantly reduces the complexity of channel estimation and symbol detection at the receiver \cite{9321356}. 

IM is a promising modulation technique for next generation wireless networks, which enjoys high spectral and energy efficiency \cite{9380189,2017Index}. In IM schemes, information bits are transmitted not only by $M$-ary signal constellations but also by the indices of building blocks, due to the extra index dimension. Many kinds of transmission entities, such as antennas \cite{5165332}, OFDM subcarriers \cite{6841601}, and radio frequency \cite{7676245}, can be used for carrying index bits without extra energy consumption. Recognizing the superiority of IM, index modulation-based OTFS (IM-OTFS) system was recently proposed to improve the bit error rate (BER) performance for high-mobility communication scenarios \cite{9129380}. The index bits are transmitted by the indices of the activated delay-Doppler resources for IM-OTFS system, where the active resource units are independent randomly selected.

However, in OTFS systems, channel delay spreads will cause severe inter-symbol interference among the resource units in the delay domain. It is necessary to take into consideration of the channel effect for OTFS system in the delay-Doppler domain when the IM scheme is applied to OTFS transmission. Unfortunately, there is no relevant work that has taken into account these factors until now. Moreover, on-the-grid channel delay or Doppler shifts are assumed in most of the existing IM-OTFS schemes, which are far beyond reality. Therefore, for OTFS systems, a sensible combination with IM is still an urgent and important research for further study.  

In this paper, we propose an efficient IM scheme for OTFS systems, named DoIM-OTFS scheme. The proposed scheme can operate with the practical rectangular pulses and work well for the practical scenarios where the channel delay and Doppler shifts do not need to land on the delay-Doppler sampling grid. Furthermore, we develop a CMP algorithm for the proposed DoIM-OTFS scheme. The CMP algorithm can effectively identify the active resource units by considering the active probability of each resource unit during the iterations. The proposed receiver algorithm can achieve desired performance with relatively low complexity. Simulation results show that our proposed DoIM-OTFS scheme provides significant performance gains compared to traditional OTFS system without IM. The proposed CMP algorithm can provide desired performance and robust to the channel uncertainties.

$Notations:$ The notations $\left ( \cdot  \right )$$^{*}$, $\left ( \cdot  \right )$$^{\rm{H}}$, and $\left \| \cdot  \right \|$ denote the conjugate, Hermitian operations, and Euclidean norm of a matrix, respectively. $\left \lfloor . \right \rfloor$ and $[\cdot ]_{m}$ denote the integer floor operator and the mod-$m$ operation, respectively. $\mathbb{C}$ and $S$ denote the ring of complex and constellation set, respectively. $C(n ,k)$ is the binomial coefficient that chooses $k$ out of $n$.  

\section{System Model}
Let us consider that a total number of $m_b$ information bits enter the DoIM-OTFS transmitter for transmission with $M$ subcarriers and $N$ symbols. The OTFS frame is split into $g$ subframes, and each subframe is divided into $\widehat{N}$ blocks along the Doppler dimension with $\widehat{M}$ delay resource units in each block. We activate the resource units based on block according to the index bits, i.e., a block of Doppler resource bins are activated simultaneously. We assume that the signal constellation symbols are normalized to have unit average power and the number of the active blocks is $\widehat{k}$. For each subframe, the OTFS index modulator processes $p=m_b/g$ bits in the delay-Doppler domain. These $p$ information bits are then divided into two parts for different purposes: the first $p_1$ bits are transferred to the index selector to decide the active resource units; the remaining $p_2$ bits are mapped to the constellation symbols and placed on active resource units. For a given $\widehat{M}$, $\widehat{N}$ and $\hat{k}$, there are C($\widehat{N}$,~$\widehat{k}$) possible index combinations of active indices and $\widehat{k}\widehat{M}$ active resource units in each subframe. The total number of index bits and constellation bits for each OTFS frame are given by $m_1=p_1g= \lfloor \log_{2}  ( {\rm C}({\widehat{N}},\widehat{k}))\rfloor g$ and $m_2=p_2g= \widehat{k}(\log_{2}  M_c)\widehat{M} g$, respectively. $M_c$ represents the modulation order.

\begin{figure}
	\centering
	\includegraphics[width=3.5in,height=2.5in]{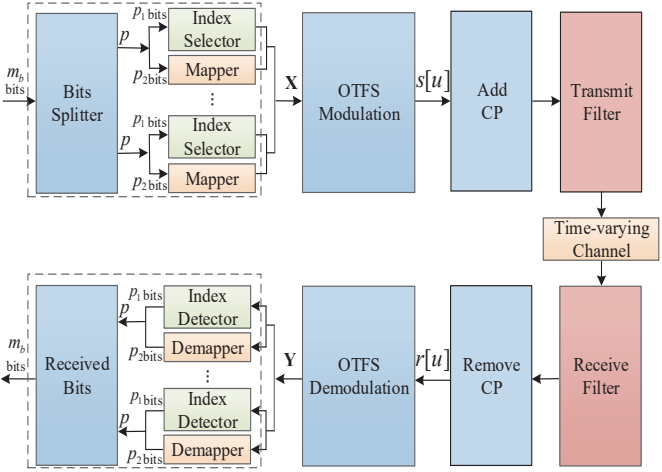}
	\caption{Transmitter and receiver structures of the DoIM-OTFS scheme.}
	\label{fig:fig3}
\end{figure}

According to the above introduced DoIM-OTFS scheme, the delay-Doppler signal $\textbf {X}\in \mathbb{C}^{M\times N}$ can be generated. Then, the signal $\textbf {X}$ is converted into the time-frequency domain by using the 2D inverse symplectic finite Fourier transform (ISFFT), 
\begin{equation}
\overline{\mathbf{X}}=\mathbf{F}_{M} \mathbf{X} \mathbf{F}_{N}^{\text{H}},
\end{equation}
where $\mathbf{F}_{M} $ and $\mathbf{F}_{N} $ denote the normalized discrete Fourier transform (DFT) matrices of size $M \times M$ and size $N \times N$, respectively. After ISFFT, the time-frequency signal $\overline{\mathbf{X}}$ is modulated through the Heisenberg transform by utilizing a transmit rectangular pulse $g_{tx}(t)$. Thus, the resulted time domain signal $\mathbf{s} \in \mathbb{C}^{M N \times 1}$ can be written as
\begin{equation}
\begin{gathered}
s[u]=\sum_{n=0}^{N-1} \sum_{m=0}^{M-1} \overline{X}[m, n] g_{t x}\left(u T_{s}-n T\right) e^{j 2 \pi m \Delta f\left(u T_{s}-n T\right)}, \\
u=0, \ldots, M N-1,\\
\end{gathered}
\end{equation}
where $\Delta f$ is the subcarrier spacing and $T_{s}=1 / M \Delta f$ denotes the symbol spaced sampling interval.

To eliminate the inter-frame interference, a cyclic prefix (CP) of length no shorter than the maximal channel delay spread is appended to the front of signal $\textbf {s}$. Then, the time domain signal $\textbf {s}$ enters the multipath fading channels after passing a transmit filter, the channel response $h[u, p]$ is characterized~as
\begin{equation}
\begin{aligned}
h[u, p] &=\sum_{i=1}^{L} h_{i} e^{j 2 \pi \nu_{i}\left(u T_{s}-p T_{s}\right)} {\rm{P}_{rc}}\left(p T_{s}-\tau_{i}\right), \\
u &=0, \ldots, M N-1 ,~ p=0,\ldots, P-1,
\end{aligned}
\end{equation}
where $h_i$ , $\tau_i$ and $v_i$ denote the channel gain, delay, and Doppler shift corresponding to the $i$-th path, respectively. $L$ represents the number of multipaths. The value of the channel tap $P$ is determined by the maximal channel delay spread and the duration of the overall filter response. ${\rm{P}_{rc}}\left(p T_{s}-\tau_{i}\right)$ is the sampled overall filter response composed of bandlimiting matched filters equipped at the transmitter and receiver. In our proposed DoIM-OTFS system, we choose a pair of root raised-cosine (RRC) filters, which are the most commonly implemented pulse shaping filters to generate an equivalent raised-cosine rolloff pulse for ${\rm{P}_{rc}}(\tau)$. Unlike the existing works, which require channel delay spread must be on the grid. Here, we relax such ideal assumption and consider that the channel delay spread do not necessarily on the grid. The Doppler frequency shift of the $i$-th path can be written as $v_{i}=\left(k_{\nu_{i}}+\beta_{\nu_{i}}\right)/{NT}$, where integer $k_{\nu_{i}}$ denotes the index of Doppler $\nu_{i}$, and $\beta_{\nu_{i}}\in(-0.5,0.5]$ represents the fractional shift from the nearest Doppler tap $k_{\nu_i}$. 



At the receiver, the channel output signal first enters a receive filter. After removing CP, the received signal can be written as
\begin{equation}
r[u]=\sum_{p=0}^{P-1} h[u, p] s\left[[u-p]_{M N}\right]+n[u], u=0, \ldots, M N-1,
\end{equation}
where $\textbf{n}=n[1],n[2],\ldots,n[MN-1]$ represents the filtered noise.

Then, the received time domain signal $\textbf{r}$ is transformed back to the time-frequency domain by Wigner transform using a rectangular pulse $g_{r x}(t)$ at the receiver, which is given by
\begin{equation}
\begin{aligned}
\overline{Y}[m, n]=& \sum_{u=0}^{M N-1} g_{r x}^{*}\left(u T_{s}-n T\right) r[u] e^{-j 2 \pi m \Delta f\left(u T_{s}-n T\right)}, \\
& m=0, \ldots, M-1 ; n=0, \ldots, N-1.
\end{aligned}
\end{equation}

Finally, the signal matrix in the time-frequency domain is processed via the symplectic finite Fourier transform (SFFT) to produce the delay-Doppler domain signal, which can be represented as
\begin{equation}
\mathbf{Y}=\mathbf{F}_{M}^{\text{H}} \overline{\mathbf{Y}} \mathbf{F}_{N}.
\end{equation}

The proposed DoIM-OTFS system structure is shown in Fig. 1. Based on the above analysis, the OTFS input-output relationship in the delay-Doppler domain can be written as~\cite{9349154}
\begin{equation}
\begin{aligned}
Y[\ell, k]=& \sum_{p=0}^{P-1} \sum_{i=1}^{L} \sum_{q=0}^{N-1} h_{i} {\rm{P}_{rc}}\left(p T_{s}-\tau_{i}\right) \gamma\left(k, \ell, p, q, k_{\nu_{i}}, \beta_{\nu_{i}}\right) \\
& \times X\left[[\ell-p]_{M},\left[k-k_{\nu_{i}}+q\right]_{N}\right]+V[\ell, k],
\end{aligned}
\label{eq:13func}
\end{equation}
where $V[\ell,k]$ denotes the delay-Doppler domain noise sample at the output of the SFFT, and
\begin{equation}
\begin{gathered}
 \hspace{-55mm}\gamma\left(k, \ell, p, q, k_{\nu_{i}}, \beta_{\nu_{i}}\right) \\
 = \begin{cases}\frac{1}{N} \xi\left(\ell, p, k_{\nu_{i}}, \beta_{\nu_{i}}\right) \theta\left(q, \beta_{\nu_{i}}\right), & p \leq \ell<M, \\
\frac{1}{N} \xi\left(\ell, p, k_{\nu_{i}}, \beta_{\nu_{i}}\right) \theta\left(q, \beta_{\nu_{i}}\right) \phi\left(k, q, k_{\nu_{i}}\right), & 0 \leq \ell<p,\end{cases} \\
\xi\left(\ell, p, k_{\nu_{i}}, \beta_{\nu_{i}}\right)=e^{j 2 \pi\left(\frac{\ell-p}{M}\right)\left(\frac{k_{\nu_{i}}+\beta_{\nu_{i}}}{N}\right)}, \\
\theta\left(q, \beta_{\nu_{i}}\right)=\frac{e^{-j 2 \pi\left(-q-\beta_{\nu_{i}}\right)}-1}{e^{-j \frac{2 \pi}{N}\left(-q-\beta_{\nu_{i}}\right)}-1}, \\
\phi\left(k, q, k_{\nu_{i}}\right)=e^{-j 2 \pi \frac{\left[k-k_{\nu_{i}}+q\right]_{N}}{N}}.
\end{gathered}
\label{eq:8func}
\end{equation}

\section{Proposed CMP Algorithm}
According to (\ref{eq:13func}), the input-output relationship of the DoIM-OTFS scheme can be vectorized as 
\begin{equation}
\mathbf{y}=\mathbf{H} \mathbf{x}+\mathbf{v},
\label{eq:15func}
\end{equation}
where $\mathbf{x}, \mathbf{y}\in \mathbb{C}^{M N \times 1}$, and $\mathbf{v}$ is the $MN \times$ $1$ noise vector. $\mathbf{H} \in \mathbb{C}^{M N \times M N}$ is a sparse matrix since the number of non-zero elements in each row and column of $\mathbf{H}$ is $Z$ due to the modulo-$N$ and modulo-$M$ operations.  
In our proposed CMP algorithm, we identify the active resource units by considering the active probability of each resource unit during the iterations. Thus, the joint maximum a posterior (MAP) probability detection rule of the transmitted signal is given by 
\begin{equation}
\widehat{\mathbf{x}}=\underset{\mathbf{x} \in \{S \cup 0\}^{NM\times 1}}{\arg \max } \operatorname{Pr}(\mathbf{x} | \mathbf{y},\mathbf{H}),
\label{eq:22func}
\end{equation}
which has a complexity exponential in $N M$. This joint MAP detection can be intractable for practical large values of $N$ and $M$. To address this problem, we calculate (\ref{eq:22func}) through the following approximation:
\begin{equation}
\begin{aligned}
{\widehat x}\left [ c \right ]&=\underset{x\in \{S \cup 0\}}{\arg \max } \operatorname{Pr}(x[c]=x | \mathbf{y},\mathbf{H})\\
&\approx \underset{x\in \{S \cup 0\}}{\arg \max }\prod_{d\in \mathcal{J}(c)}{\rm{ Pr }}\left ( y[d] | x[c]=x,\mathbf{H}\right ).
\end{aligned}
\label{eq:25func}
\end{equation}

For the proposed CMP algorithm, we aim to detect each element $x[c]$ of the transmit signal vector $\mathbf{x}$ based on (\ref{eq:15func}). Let $\mathcal{I}(d)$ and $\mathcal{J}(c)$ denote the sets of indexes with nonzero elements in the $d$-th row and $c$-th column of $\mathbf{H}$, where $d=1, \ldots,  N M$ and $c=1, \ldots, N M$, respectively. Each observation node $y[d]$ is connected to the set of $Z$ variable nodes $\{x[c], c \in \mathcal{I}(d)\}$ while each variable node $x[c]$ is connected to the set of $Z$ observation nodes $\{y[d], d \in \mathcal{J}(c)\}$. We assume the components of $\mathbf{y}$ are approximately independent for a given $x[c]$ owing to the sparsity of $\mathbf{H}$, and any transmitted symbol $x\in \{S \cup 0\}$ with equal probability.. The variable node $x[c]$ is isolated from the other interference terms for each $y[d]$, which has an easily computable mean and variance. 
Based on (\ref{eq:25func}), we interpret the system model as a sparsely connected factor graph with two types of nodes: $(i)~ MN $ observation nodes corresponding to the elements of $\mathbf{y}$, $(ii)~MN$ variable nodes corresponding to the elements of $\mathbf{x}$. For the proposed CMP algorithm, its detailed steps in iteration $n_{iter}$ are described below. 


$\textbf{1) From observation node}~ ~y[d] ~~\textbf{to variable nodes}~~x[c],$\\$~c\in \mathcal{I} (d)$:
At each observation node, we calculate the extrinsic messages for each connected variable node according to the sparse channel model, and prior information from other connected variable nodes. The mean $\mu_{d, c}^{n_{iter}}$ and variance $(\sigma_{d, c}^{n_{iter}})^{2}$ of the interference are approximately modeled as a Gaussian random variable $\zeta_{d, c}^{n_{iter}}$. Thus, the received signal $y[d]$ can be written as
\begin{equation}
y[d]=x[c] H[d, c]+\underbrace{\sum_{e \in \mathcal{I}(d), e \neq c} x[e] H[d, e]+v[d]}_{\zeta_{d, c}^{n_{iter}}},
\label{eq:23func}
\end{equation}
with
\begin{equation}
\mu_{d, c}^{n_{iter}}=\sum_{e \in \mathcal{I}(d), e \neq c} H[d, e]\sum_{x\in \{S \cup 0\}} p_{e, d}^{n_{iter}-1}\left(x\right) x ,
\label{eq:27func}
\end{equation}
and
\begin{equation}
\begin{aligned}
(\sigma_{d, c}^{n_{iter}})^{2}=& \sum_{e \in \mathcal{I}(d), e \neq c}\left(\sum_{x\in \{S \cup 0\}} p_{e, d}^{n_{iter}-1}\left(x\right)\left|x\right|^{2}|H[d, e]|^{2}\right.\\
&\left.-\left|\sum_{x\in \{S \cup 0\}} p_{e, d}^{n_{iter}-1}\left(x\right) x H[d, e]\right|^{2}\right)+\sigma^{2} ,
\label{eq:28func}
\end{aligned}
\end{equation}
where $\sigma^{2}=\sigma _{N}^{2}\int _{\mu }{\rm{P}_{rrc}^{2}}(\mu )d\mu $ is the variance of the colored Gaussian noise. ${\rm{P}_{rrc}}(\mu )$ denotes the RRC rolloff receive filter and $\sigma _{N}^{2}$ is the variance of the AWGN at the receiver input \cite{9349154}. Therefore, the probability estimate of $x[c]$ passed from observation node $y[d]$ to variable node $x[c]$ is given by
\begin{equation}
\begin{aligned}
v_{d,c}^{n_{iter}}(x)&\overset{\Delta }{=}{\rm{ Pr }}(x[c]=x|y[d])\\
&\propto \exp \left (-\frac{ \left | y[d]-\mu _{d,c}^{n_{iter}}-H[d,c]x \right |^{2}}{(\sigma  _{d,c}^{n_{iter}})^2}\right ).
\end{aligned}
\label{eq:26func}
\end{equation}

$\textbf {2) From variable node}~~x[c]~~\textbf {to observation nodes}~~y[d],\\d\in \mathcal{J} (c)$: The posterior probability of the elements $\mathbf{x}$ passed from variable node $x[e]$ to observation node $y[d]$ is denoted by $\textbf{p}_{e,d}^{n_{iter}}$. At each variable node, the extrinsic information for each connected observation node is generated from prior messages collected from other observation nodes. The message from variable node $x[c]$ to observation node $y[d]$ is the probability mass function (pmf) $\textbf{p}_{c,d}^{n_{iter}}$ with elements 
\begin{equation}
\begin{aligned}
p_{c,d}^{n_{iter}}\left ( x \right )=\Delta \cdot \widetilde{p}_{c,d}^{n_{iter}}(x)+\left ( 1-\Delta  \right )\cdot p_{c,d}^{n_{iter-1}}\left ( x \right ),
\label{eq:36func}
\end{aligned}
\end{equation}
where $\Delta \in (0,1]$ is the message damping factor used to improve the system performance by controlling the convergence rate, and
 \begin{equation}
\begin{aligned}
\widetilde p_{c,d}^{n_{iter}}(x)&\propto \prod_{e\in \mathcal{J}(c),e\neq d}{\rm{ Pr }}\left ( y[e] \mid x[c]=x,\mathbf{H}\right )\\
&=\prod_{e\in \mathcal{J}(c),e\neq d}\frac{v_{e,c}^{n_{iter}}(x)}{\underset{x\in \{S \cup 0\}}{\sum }v_{e,c}^{n_{iter}}(x)}.
\end{aligned}
\end{equation}

$\textbf{3) Convergence ~indicator }$: We calculate the convergence indicator $\eta ^{n_{iter}}$ for some small $\varrho$ as 
\begin{equation}
\begin{aligned}
\eta ^{n_{iter}}=\frac{1}{MN}\sum_{c=1}^{MN}\mathbb{I}\left ( \underset{x\in \{S \cup 0\}}{{\text{max}}} ~p_{c}^{n_{iter}}\left ( x \right )\geq 1-\varrho \right ),
\label{eq:37func}
\end{aligned}
\end{equation}
where $\mathbb{I}$ denotes indicator function. The posterior probability for each element of the transmit symbol is given by 
\begin{equation}
\begin{aligned}
p_{c}^{n_{iter}}\left ( x\right )=\prod_{e\in \mathcal{J}(c)}\frac{v_{e,c}^{n_{iter}}(x)}{\underset{x\in \{S \cup 0\}}{\sum }v_{e,c}^{n_{iter}}(x)}, ~\forall x\in\{S \cup 0\}.
\end{aligned}
\end{equation}

$\textbf{4) Update~ criteria}$: If $\eta ^{n_{iter}}>\eta ^{n_{iter}-1}$, the probability of the transmitted symbols is updated only when the current iteration provides a better solution than the previous one,
\begin{equation}
\begin{aligned}
\overline{\mathbf{ p }}_{c}=\mathbf{ p }_{c}^{n_{iter}}, c=1,\ldots ,MN.
\end{aligned}
\end{equation} 

In this algorithm, $\textbf{v}_{d,c}^{n_{iter}}$ passes from observation node $y[d]$ to variable node~$x[c]$; $\textbf{p}_{d,c}^{n_{iter}}$ passes from variable node $x[c]$ to observation node $y[d]$. All of the messages are exchanged between these two nodes.

$\textbf{5) Stopping~ criteria}$: The CMP algorithm stops when $\eta ^{n_{iter}}=1$ or the maximum number of iterations $n_{iter}^{max}$ is reached. 

Once the stopping criteria is satisfied, we can obtain the log-likelihood ratio (LLR) of each resource unit, and then, average the LLRs of all resource units in each block. According to the value of $\widehat{k}$, the active resource units can be determined by choosing the blocks of the corresponding $\widehat{k}$ largest average LLRs. Finally, we make the decisions of the transmitted symbols ${\widehat x}\left [ c \right ]$ for the active resource units,~as
 \begin{equation}
\begin{aligned}
{\widehat x}\left [ c \right ]=\underset{x\in S} {\text{argmax}} \overline{p}_{c}\left ( x \right ), ~c \in \mathbb{A},
\label{eq:41func}
\end{aligned}
\end{equation}
where $\mathbb{A}$ denotes the set of active resource units.

\section{Simulation Results}
In this section, we study the BER performance of the proposed DoIM-OTFS scheme with CMP algorithm, and provide a comparison with the classical MP algorithm in OTFS systems without IM. We consider the carrier frequency is 4 GHz and subcarrier spacing is $\Delta f$ = 15 kHz. The Doppler shift of $i$-th channel path is generated by $v_i=v_{max}\text{cos}(\theta_i)$, where $v_{max}$ denotes the maximum Doppler shift with parameter $-\pi\leq\theta_i\leq\pi$. Moreover, the RRC rolloff factor is set to 0.4 at the transmitter and receiver. The time slots $N=32$ and subcarriers $M=64$ are considered in the time-frequency domain and the number of multipaths is set to $L=4$. Without loss of generality, we choose $\Delta$ = 0.4 and $\varrho=0.1$. Unless otherwise mentioned, the number of delay bins, Doppler bins and active blocks in each subframe are set to $\widehat{M}=4$, $\widehat{N}=4$ and $\widehat{k}=1$, respectively, and the user velocity is 300 Kmph.

\begin{figure}
	\center
	\includegraphics[width=3.2in,height=2.3in]{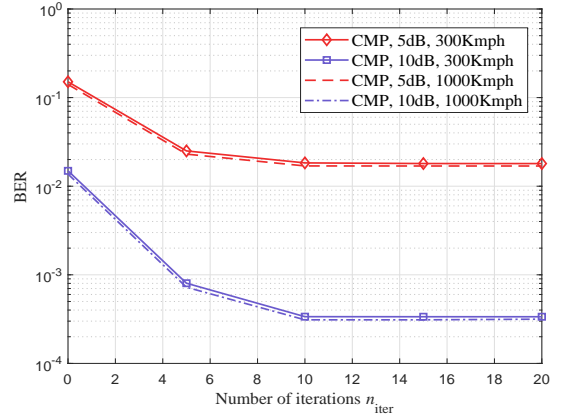}
	\caption{BER convergence of the proposed CMP algorithm at SNR = 5~dB and 10~dB under different user velocities.}
	\label{fig:fig8}
\end{figure}

In Fig.~\ref{fig:fig8}, we illustrate the convergence speed of the proposed CMP algorithm for different SNRs. The user velocities are set to 300Kmph and 1000Kmph, respectively. As shown in the Fig. 2, the proposed CMP algorithms under low SNR exhibit a slightly faster convergence speed than that of high SNR. Moreover, the error performance of CMP algorithm with different user velocities is similar as no additional diversity can be achieved with the increase of velocity. At an SNR of 5~dB, the CMP algorithm converges after 8 iterations on average. While for high SNR of 10~dB, the CMP algorithm converges in 10 iterations. Based on the above analysis, we take the number of iterations to be 10 for the following simulation tests.

\begin{figure}
	\center
	\includegraphics[width=3.2in,height=2.3in]{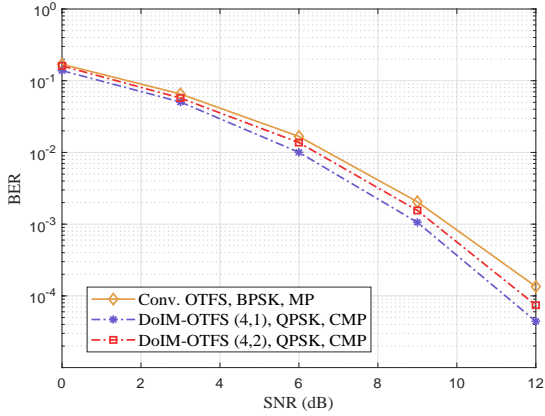}
	\caption{BER performance comparison between the proposed DoIM-OTFS scheme and the existing OTFS system without IM.}
	\label{fig:fig9}
\end{figure}

In Fig.~\ref{fig:fig9}, we compare the BER performance of the proposed DoIM-OTFS scheme and the traditional OTFS system without IM. For brevity, we will refer to “DoIM-OTFS $({\widehat{N}},\widehat{k})$” as the DoIM-OTFS scheme with $\widehat{k}$ out of $\widehat{N}$ blocks are activated. From Fig.~\ref{fig:fig9}, it can be observed that the BER curve of the proposed CMP algorithm for the DoIM-OTFS scheme is better than that of the conventional OTFS system without IM. Specifically, the CMP algorithm with the ``DoIM-OTFS (4,1)'' scheme shows an SNR gain of nearly 1~dB over the existing OTFS system without IM. 


\begin{figure}
	\center
	\includegraphics[width=3.2in,height=2.3in]{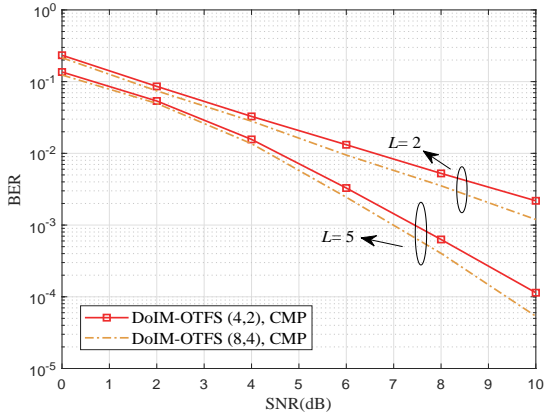}
	\caption{BER performance comparison between different active indices of the proposed DoIM-OTFS system under different propagation paths.}
	\label{fig:fig10}
\end{figure}

Fig.~\ref{fig:fig10} gives the comparison results for DoIM-OTFS scheme with parameters ``$\widehat{N}=\widehat{M}=4,~\widehat{k}=2$" and ``$\widehat{N}=8,~\widehat{M}=4,~\widehat{k}=4$" under different number of multipaths. The spectral efficiency of ``(4,~2)" and ``(8,~4)" are 1.125 and 1.188bps/Hz, respectively. As seen from Fig. 4, we can observe that the error performance of the DoIM-OTFS scheme degrades with the decrease of $\widehat{N}$ due to severe inter-block interference. Furthermore, the DoIM-OTFS system of $L=5$ can provide superior performance than that of $L=2$ since the number of independent resolvable paths is higher for large value of $L$. As a result, better diversity gain can be achieved.


Finally, the BER performance of the proposed CMP algorithm is tested in terms of imperfect CSI. Here, we characterize the CSI errors by adopting the following model
\cite{9354639}

\begin{figure}
	\center
	\includegraphics[width=3.2in,height=2.3in]{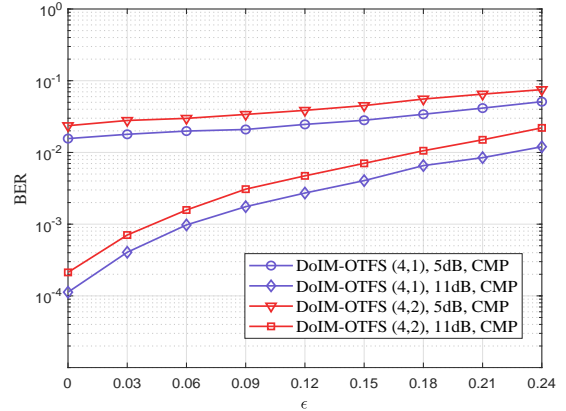}
	\caption{BER performance of the DoIM-OTFS scheme with imperfect CSI.}
	\label{fig:fig12}
\end{figure}

\hspace{20mm}$h_{i}=\tilde{h}_{i}+\Delta h_{i},\left \| \Delta h_{i} \right \|\leq \epsilon_{ h_{i}}$,

\hspace{20mm}$v_{i}=\tilde{v}_{i}+\Delta v_{i},\left \| \Delta v_{i} \right \|\leq \epsilon_{ v_{i}}$,

\hspace{20mm}$\tau_{i}=\tilde{\tau}_{i}+\Delta \tau_{i},\left \| \Delta \tau_{i} \right \|\leq \epsilon_{ \tau_{i}}$,

\hspace{-3mm}where $\tilde{h}_{i}$, $\tilde{v}_{i}$ and $\tilde{\tau}_{i}$ denote the estimated values of $h_{i}$, $v_{i}$, and $\tau_{i}$, respectively. $\Delta h_{i}$, $\Delta v_{i}$, and $\Delta \tau_{i}$ are the corresponding channel estimate errors. The norm of $\Delta h_{i}$, $\Delta v_{i}$, and $\Delta \tau_{i}$ should not exceed the given values of $\epsilon_{ h_{i}}$, $\epsilon_{ v_{i}}$ and $\epsilon_{ \tau_{i}}$, respectively. Here, we assume $\epsilon_{ h_{i}}=\epsilon\left \| \tilde{h}_{i} \right \|$, $\epsilon_{ v_{i}}=\epsilon\left \| \tilde{v}_{i} \right \|$ and $\epsilon_{ \tau_{i}}=\epsilon\left \| \tilde{\tau}_{i} \right \|$. From Fig.~\ref{fig:fig12}, we can observe that the proposed CMP algorithm only suffers from mild performance loss for the modest values of channel uncertainty $\epsilon$. With the increase of the level of channel uncertainty, the rapid degradation in BER performance does not appear, which verifies the robustness of our proposed CMP algorithm.

\section{Conclusion}
In this paper, we proposed a DoIM-OTFS scheme to further improve the BER performance of the conventional OTFS systems. Our proposed IM scheme took into account the channel effects of OTFS system in the delay-Doppler domain, and robust to the doubly-selective fading channels for high mobility communications. Moreover, we developed a low complexity CMP algorithm for our proposed DoIM-OTFS system. Simulation results demonstrated that our proposed DoIM-OTFS scheme outperformed the traditional OTFS system without IM. The proposed algorithm can achieve desired performance and robust to the imperfect CSI scenario.

\bibliographystyle{IEEEtran}
\bibliography{reference}

\end{document}